\documentclass[[12pt,a4paper,preprint,preprintnumbers,amsmath,amssymb,nofootinbib]{revtex4}
\usepackage{graphicx}
\usepackage{amsmath}
\usepackage{amssymb}
\usepackage{bbold}
\usepackage{type1cm}
\usepackage{rotating}
\usepackage{tabularx}
\usepackage{pdflscape}

\usepackage{dsfont}
\usepackage{overpic}
\usepackage{natbib}
\usepackage{slashed}
\usepackage{subfigure}
\usepackage{bbm}

\begin{document}
\title{Spatial interpretation of ``compositeness'' for finite-range potentials}

\author{Peter C.~Bruns}
\affiliation{Nuclear Physics Institute, 25068 \v{R}e\v{z}, Czech Republic }
\date{\today}
\begin{abstract}
We discuss the relation between the ``compositeness'' of an s-wave bound state, as derived from a related partial wave scattering amplitude, and the corresponding spatial probability densities, for the case of spherically symmetric, energy-independent finite-range potentials in non-relativistic quantum mechanics. We find that in this simple case ``compositeness'' is a measure for the probability to find the constituents separated by a distance greater than the interaction range.
\end{abstract}

\maketitle

\section{Introduction}
\label{sec:Intro}

In the past few years, the concept of ``compositeness'' of hadronic bound states and resonances \cite{Sekihara:2014kya,Hyodo:2008xr,Gamermann:2009uq,Aceti:2012dd,Hyodo:2013nka,Nagahiro:2014mba,Guo:2015daa,Molina:2016pbg,Sekihara:2016xnq,Oller:2017alp} has played a prominent role in the discussion on the ``nature'' of hadrons (hadronic molecules vs. quark-antiquark or three-quark states, etc., see \cite{Guo:2017jvc} for a recent review). The debate on this topic can be traced back to the time when  ``so many new strongly interacting particles were being discovered at accelerators like the Bevatron that we had to give up on identifying which of them were elementary particles, whose fields would appear in the Lagrangian, and we even began to doubt if we knew what was meant by a particle being elementary" \cite{Weinberg:2018apv}, and when subsequently quantitative methods were developed \cite{Weinberg:1962hj,Weinberg:1965zz,Vaughn:1961poz,Zeldovich:1961b,Salam:1962ap} to assess the various ``components'' in a given hadronic state (see also the more recent work in \cite{Baru:2003qq,Baru:2010ww,Pelaez:2003dy}). Considering a hadronic bound-state or resonance $H$ that couples to some hadronic two-body state, the ``compositeness'' $\mathcal{C}$ is basically given by $1-\mathcal{Z}$, where $\mathcal{Z}$ (the field renormalization constant pertaining to $H$) is sometimes interpreted as the probability to find a bare ``elementary'' state in the state $H$ (or, more generally, anything else than the composite two-body state)\footnote{Strictly speaking, in no measurement does one ever ``find a bare state in the bound state'', but one might find, e.g. in a measurement of form factors of a bound state, a ``hard core'', i.e. a structure that is noteably smaller than the typical size expected for a two-body bound state with a given interaction.}. For $\mathcal{Z}=0$, $H$ is considered a ``purely composite'' state. \\
In the present work, we shall follow the more operational prescription  of deriving the ``compositeness'' from on-shell partial wave amplitudes (see Eqs.~(\ref{eq:Tellform})-(\ref{eq:defredcomp}) below), which is used in practical applications in hadron physics, where the wave functions are not known beforehand, but parameterizations of partial wave amplitudes are readily at hand. This will allow for a largely model-independent interpretation of the extracted ``compositeness''. \\
We consider spherically symmetric finite-range potentials $V$, strictly vanishing beyond some distance $d$ ($0<d<\infty$), and show that in this case, the ``compositeness'' of an s-wave bound state (in the sense just alluded to) is closely related to another probability, namely the one to find the two (point) particles of the two-body state separated by a distance greater than the range of their interaction. Although this is pretty straightforward to show, we didn't find an explicit statement to that effect in the literature. We assume non-relativistic dynamics, and treat the two-particle system as one particle of reduced mass $\mu$ in a fixed potential centered at the origin.
Let us emphasize that we stay in the realm of standard non-relativistic quantum mechanics here, and do {\em not\,} consider potentials $V$ with an explicit energy dependence. For energy-dependent potentials, the bilinear form used to determine scalar products (normalizations, expectation values) has to be modified \cite{Lepage:1977gd,Sazdjian:1986qn,Mares:2004,Lombard:2007zz}; otherwise, wave functions for different energy eigenvalues are not guaranteed to be orthogonal to each other, and properly normalized, any more (and one would have to search for generalizations of completeness relations). Still, the {\em effective\,} potentials encoded in the on-shell partial wave amplitudes, from which the ``compositeness'' is usually inferred in practice, will in general be energy-dependent, see e.g. Eqs.~(\ref{eq:Tellform}), (\ref{eq:E00}) below. As we will see, this leads to values for the ``compositeness'' different from one, while the bound-state wave functions are normalized in the standard way for energy-independent potentials. 


\section{LSE and asymptotic wave functions}
\label{sec:LSE}

We start with the Lippmann-Schwinger equation (LSE) for the off-shell scattering amplitude $\mathcal{T}(\vec{q}\,',\vec{q};E)$ of fixed energy $E$, with $\vec{q},\,\vec{q}\,'$ the off-shell three-momenta of the incoming and outgoing particle, respectively:
\begin{equation}
  \mathcal{T}(\vec{q}\,',\vec{q};E) = V(\vec{q}\,',\vec{q}) + \int d^3l\frac{\mathcal{T}(\vec{q}\,',\vec{l};E)V(\vec{l},\vec{q})}{E-\frac{|\vec{l}|^2}{2\mu}}\,,\quad V(\vec{q}\,',\vec{q}) = \int\frac{d^3x}{(2\pi)^3}\,e^{-i(\vec{q}\,'-\vec{q})\cdot\vec{x}}V(\vec{x})\,.\label{eq:LSE}
\end{equation}
See e.g. the textbook \cite{Taylor}. It is understood that the physical $E$-axis is to be approached from the upper complex plane, $E\rightarrow E+i\epsilon$. Formally, the solution can be written as
\begin{equation}\label{eq:Tsolform}
\mathcal{T}(\vec{q}\,',\vec{q};E) = V(\vec{q}\,',\vec{q}) + \sum_{n,\ell,m}\frac{\left(\frac{|\vec{q}\,'|^2}{2\mu}-E_{n}\right)\tilde{\psi}_{n\ell m}(\vec{q}\,')\tilde{\psi}^{\ast}_{n\ell m}(\vec{q})\left(\frac{|\vec{q}|^2}{2\mu}-E_{n}\right)}{(2\pi)^3(E-E_{n})}\,,
\end{equation}
with a complete set of momentum-space eigenfunctions for the Hamiltonian with potential $V$, normalized as 
\begin{center}
  \hspace{-1.2cm}$\int\frac{d^3p}{(2\pi)^3}\tilde{\psi}^{\ast}_{n'\ell' m'}(\vec{p})\tilde{\psi}_{n\ell m}(\vec{p}) = \delta_{n'n}\delta_{\ell'\ell}\delta_{m'm}$ \quad for the discrete part of the spectrum, and \\
  $\int\frac{d^3p}{(2\pi)^3}\tilde{\psi}^{\ast}_{n'\ell'm'}(\vec{p})\tilde{\psi}_{n\ell m}(\vec{p}) = \delta(E_{n'}-E_{n})\delta_{\ell'\ell}\delta_{m'm}$ \quad for the continuous part of the spectrum. 
\end{center}
Completeness here means $\sum_{n\ell m}\tilde{\psi}_{n\ell m}(\vec{p}\,')\tilde{\psi}^{\ast}_{n\ell m}(\vec{p})=(2\pi)^3\delta^{3}(\vec{p}\,'-\vec{p})$. For the continuum part of the spectrum, the summation over the eigenvalues $E_{n}$ is of course to be replaced by an integration $\sim\int dE_{n}\,$. To see that this form of the solution makes sense, just plug it as an ansatz in Eq.~(\ref{eq:LSE}), and use the fact that the wave functions solve the Schr\"odinger equation
\begin{equation}\label{eq:schreq}
\left(\frac{p^2}{2\mu}-E_{n}\right)\tilde{\psi}_{n\ell m}(\vec{p})+\int d^3l\,V(\vec{p},\vec{l}\,)\tilde{\psi}_{n\ell m}(\vec{l})=0\,,\qquad p:=|\vec{p}|\,,
\end{equation}
together with the completeness relation. As we consider spherically symmetric potentials $V(\vec{x})$, we can write $\vec{x}$ in spherical coordinates $r,\theta_{x},\phi_{x}$, and split off a spherical harmonic $\mathcal{Y}_{\ell m}(\theta_{p},\phi_{p})$ in all wave functions $\tilde{\psi}_{n\ell m}(\vec{p})$, which contains all the angular dependence. \\
For almost all practical purposes, one is mainly interested in the on-shell partial wave amplitudes following from $\mathcal{T}(\vec{q}\,',\vec{q};E)$. Denoting $k:=+\sqrt{2\mu E}$ (selecting the square root with the positive imaginary part for complex $E$), these partial wave amplitudes are
\begin{equation}
\mathcal{T}_{\ell}(E):=\frac{1}{2}\int_{-1}^{1}dz\,\mathcal{P}_{\ell}(z)\,\mathrm{lim}_{|\vec{q}\,'|,|\vec{q}|\rightarrow k}\mathcal{T}(\vec{q}\,',\vec{q};E)\,,\qquad f_{\ell}(k):=-(2\pi)^2\mu\mathcal{T}_{\ell}(E)\,.
\end{equation}
Here $z$ is the cosine of the scattering angle (the angle between $\vec{q}\,'$ and $\vec{q}$), and $\mathcal{P}_{\ell}(z)$ denote the usual Legendre polynomials. The conventionally normalized partial wave amplitudes $f_{\ell}(k)$ obey the unitarity requirement $\mathrm{Im}\,(f_{\ell}(k))^{-1}=-k\,$ for real $E>0$.
Now, the spatial s-wave bound-state wave functions for energy $E=E_{B}<0$ pertaining to the finite range potential behave as 
\begin{equation}\label{eq:psiB00asymp}
\psi_{B00}(\vec{x}) = \mathcal{N}_{B}\frac{e^{-\kappa_{B}r}}{r}\mathcal{Y}_{00}(\theta_{x},\phi_{x})\,\quad\mathrm{for}\quad r>d\,,
\end{equation}
i.e., outside the range of the potential, while the continuum wave functions $\psi_{E00}(\vec{x})$ behave as $\sim (e^{ikr}kf_{0}(k)+\sin(kr))/r$ in that region. In Eq.~(\ref{eq:psiB00asymp}), we denote $k_{B}=k(E\rightarrow E_{B})=i\kappa_{B}$, or $\kappa_{B}=+\sqrt{-2\mu E_{B}}$. The prefactor $\mathcal{N}_{B}$ is fixed by the normalization $\int d^{3}x\,|\psi_{B00}(\vec{x})|^2\overset{!}{=}1\,$ (choosing the phase of the wave function such that it is real). For the probability to find the bound particle outside the range of the potential, we thus obtain
\begin{equation}\label{eq:Prd}
P(r>d) = \frac{\mathcal{N}_{B}^2}{2\kappa_{B}}e^{-2\kappa_{B}d}\,.
\end{equation}
Let us now study the residue of $f_{\ell=0}(k)$ at the bound-state pole. From the above equations, there can only be a non-vanishing residue if
\begin{equation}\label{eq:B00tilde}
\tilde{\psi}_{B00}(\vec{p}) = \int d^3x\,e^{-i\vec{p}\cdot\vec{x}}\psi_{B00}(\vec{x})
\end{equation}
has a pole $\sim (p^2-k_{B}^2)^{-1}$. For a sufficiently regular potential, such a pole can only be generated by the long-distance part of the integral in Eq.~(\ref{eq:B00tilde}), since the Fourier integral of a sufficiently regular integrand over a finite region (e.g. $r<d$) cannot produce such a strong singularity (Paley-Wiener theorem). So we can in fact calculate the pole term from the asymptotic form of Eq.~(\ref{eq:psiB00asymp}),
\begin{equation}
\tilde{\psi}_{B00}(\vec{p}) = \frac{\sqrt{4\pi}\mathcal{N}_{B}}{p^2-k_{B}^2} + \mathrm{regular\,\,terms}\,,
\end{equation}
and the residue of the s-wave amplitude at the pole $E=E_{B}$ then follows straightforwardly,
\begin{equation}\label{eq:resf0}
\mathrm{Res}\,f_{0}(k_{B}) = -\frac{\mathcal{N}_{B}^2}{2\mu}\,.
\end{equation}


\section{Extraction of ``compositeness''}
\label{sec:comp}

The partial wave amplitudes of a solution to a LSE are often written in the form
\begin{equation}\label{eq:Tellform}
\mathcal{T}_{\ell}(E) = \left[W_{\ell,\mathrm{eff}}^{-1}(E)-G_{\mathrm{reg}}(E)\right]^{-1}\,,
\end{equation}
where $W_{\ell,\mathrm{eff}}(E)$ plays the role of an effective partial-wave potential, and $G_{\mathrm{reg}}(E)$ is a regularized version of the ``loop'' integral $\int\frac{d^3l}{E-\frac{l^2}{2\mu}}$\,. Assume that the above amplitude has a simple bound-state pole at $E_{B}$, with residue $\mathrm{Res}\,\mathcal{T}_{\ell}(E_{B})$. Then the second term in the splitting
\begin{equation}\label{eq:composplit}
1 = \mathrm{Res}\,\mathcal{T}_{\ell}(E_{B})\frac{dW_{\ell,\mathrm{eff}}^{-1}}{dE}\biggr|_{E_{B}} + \mathrm{Res}\,\mathcal{T}_{\ell}(E_{B})\left(-\frac{dG_{\mathrm{reg}}}{dE}\biggr|_{E_{B}}\right)
\end{equation}
gives the ``compositeness'' of the bound state $B$, while the remaining first term is sometimes called ``elementariness'' (see e.g. \cite{Sekihara:2014kya})\footnote{This prescription is commonly used in the context of separable potential models with an explicit energy dependence. Our aim here, however, is to make sense of the quantities defined above in a model-independent way, avoiding a definition via wave function normalization factors. See also the careful discussion in \cite{Nagahiro:2014mba,Sekihara:2016xnq}.}. The typical feature of the (regularized) loop integral is that it has a branch point at the threshold energy ($E=0$), and an imaginary part fixed by unitarity, $\mathrm{Im}\,G_{\mathrm{reg}}(E)=-(2\pi)^2\mu k$ for $E>0$. The real part depends on the regularization used in the computation\footnote{Employing dimensional regularization, one obtains $\,G_{\mathrm{d.reg}}=-2\mu\pi^{\frac{D}{2}}\Gamma\left(1-\frac{D}{2}\right)(-ik)^{D-2}\,$; in our case, $D\rightarrow 3$.}, and is typically regular in $E$, at least below some chosen cutoff scale. It is clear from Eqs.~(\ref{eq:Tellform}), (\ref{eq:composplit}) that an energy-independent real constant in $G_{\mathrm{reg}}$ can always be absorbed in $W_{\ell,\mathrm{eff}}^{-1}$, without affecting the splitting in Eq.~(\ref{eq:composplit}). To get rid of the residual model dependence, we also absorb the possible remaining real regular terms in $E$ in the effective potential, and define our (``reduced'') compositeness as
\begin{equation}\label{eq:defredcomp}
\mathcal{C}_{B}^{\ell} := \mathrm{Res}\,\mathcal{T}_{\ell}(E_{B})(2\pi)^2\mu i\frac{dk}{dE}\biggr|_{E_{B}} = -\frac{\mu}{\kappa_{B}}\mathrm{Res}\,f_{\ell}(k_{B})\,.
\end{equation}
We expect that the term $\sim\frac{dk}{dE}\sim E^{-1/2}$ dominates the full compositeness for small energies, at least for a reasonable regularization, and so our redefinition should only induce a moderate modification in this case. - For the s-wave bound state discussed in the previous section, we thus get
\begin{equation}
\mathcal{C}_{B}^{0} = \frac{\mathcal{N}_{B}^2}{2\kappa_{B}} \approx P(r>d)\quad \mathrm{for}\quad \kappa_{B}d\ll 1\,.
\end{equation}
That is, up to the exponential factor appearing in Eq.~(\ref{eq:Prd}), which is $\approx 1$ for a weakly bound particle in a short-range potential, the ``reduced compositeness'' $\mathcal{C}_{B}^{0}$ gives the probability to find the bound particle outside the range of the potential, $\,P(r>d)=\mathcal{C}_{B}^{0}e^{-2\kappa_{B}d}\,$.

\section{Example: Spherical well}
\label{sec:sphwell}
Let us verify the results of the foregoing sections with a concrete example, the spherical well potential $V(\vec{x})=V_{0}\theta(d-r)$, where $V_{0}$ is a real constant, $r=|\vec{x}|$, and $\theta(\cdot)$ is the usual Heaviside step function. The pertinent wave functions and partial wave amplitudes can be found with standard methods, see e.g. Chapter 11 of \cite{Taylor}. \\
\underline{Bound-state wave function for $\ell=0$ ($V_{0}<E_{B}<0$):}\,
\begin{eqnarray}
  \psi_{B00}(\vec{x}) &=& \frac{\mathcal{N}_{B}}{r}\left(\theta(d-r)e^{-\kappa_{B}d}\left(\frac{\sin\xi_{B}r}{\sin\xi_{B}d}\right)+\theta(r-d)e^{-\kappa_{B}r}\right)\mathcal{Y}_{00}(\theta_{x},\phi_{x})\,,\label{eq:B001}\\
  \kappa_{B} &=& \sqrt{-2\mu E_{B}}\,,\qquad \xi_{B} = \sqrt{-2\mu V_{0}-\kappa_{B}^2}\,,\qquad \tan(\xi_{B}d) = -\frac{\xi_{B}}{\kappa_{B}}\,,\label{eq:B002}\\
  \mathcal{N}_{B}^{2} &=& \frac{2\kappa_{B}\xi_{B}^{2}e^{2\kappa_{B}d}}{(1+\kappa_{B}d)(\kappa_{B}^2+\xi_{B}^2)}\,.\label{eq:B003}
\end{eqnarray}
The corresponding momentum-space wave function is (again denoting $p:=\sqrt{\vec{p}\cdot\vec{p}}\,$)
\begin{equation}\label{eq:B00tilde_well}
  \tilde{\psi}_{B00}(\vec{p}) = \sqrt{4\pi}\mathcal{N}_{B}(\kappa_{B}^2+\xi_{B}^2)e^{-\kappa_{B}d}\left(\frac{\kappa_{B}\sin(pd)+p\cos(pd)}{p(\xi_{B}^2-p^2)(p^2+\kappa_{B}^2)}\right) \,.
\end{equation}
We point out that $\tilde{\psi}_{B00}(\vec{p})$ is nonsingular at $p=0$, and also at $p=\pm\xi_{B}$, due to the bound-state pole condition in Eq.~(\ref{eq:B002}). The pole at $k_{B}=i\kappa_{B}=\pm p$, however, is due to the long-distance part of the Fourier integral in Eq.~(\ref{eq:B00tilde}).\\
\underline{Continuum wave function for $\ell=0$ ($E>0,\,k=+\sqrt{2\mu E}$):}\,
\begin{eqnarray}
  \psi_{E00}(\vec{x}) &=& \sqrt{\frac{2\mu}{\pi k}}\biggl(\theta(d-r)\frac{k\sin(\xi r)e^{-ikd}}{\xi\cos(\xi d)-ik\sin(\xi d)} + \theta(r-d)\left(e^{ikr}kf_{0}(k)+\sin(kr)\right)\biggr)\frac{\mathcal{Y}_{00}(\theta_{x},\phi_{x})}{r} \nonumber \,, \\
  f_{0}(k) &=& \frac{k\sin(\xi d)\cos(kd)-\xi\cos(\xi d)\sin(kd)}{ke^{ikd}(\xi\cos(\xi d)-ik\sin(\xi d))} \equiv \left[K_{0}^{-1}-ik\right]^{-1}\,,\qquad \xi=\sqrt{k^2-2\mu V_{0}}\,,\label{eq:E00}\\
  K_{0} &=& \frac{\,k\sin(\xi d) \cos(kd)-\xi\cos(\xi d)\sin(kd)}{k(k\sin(\xi d)\sin(kd)+\xi\cos(\xi d)\cos(kd))}\,.\nonumber
\end{eqnarray}
Note that $K_{0}$ is even in $k$ and in $\xi$, so it is real for real $E$, and does not possess branch cuts (but it can have poles for positive energies). It plays a similar role as the effective potential $W_{\ell,\mathrm{eff}}(E)$ in Eq.~(\ref{eq:Tellform}).
- There are now two ways available to compute the residue of $f_{0}$ at the bound-state pole $E_{B}$, either from Eq.~(\ref{eq:B00tilde_well}), relying on the representation of the relevant pole term in Eq.~(\ref{eq:Tsolform}), or directly from the (asymptotic) form of the continuum solution in Eq.~(\ref{eq:E00}), where $f_{0}$ appears explicitly. We find that both results agree, and indeed confirm Eq.~(\ref{eq:resf0}). \\
The associated compositeness, as defined in Eq.~(\ref{eq:defredcomp}), is equal to one for $E_{B}\rightarrow 0$, and approaches zero when $\frac{E_{B}}{V_{0}}\rightarrow 1$ for $V_{0}\not=0$. A probabilistic interpretation of $\mathcal{C}_{B}^{0}$ can only be assumed if $\mathcal{C}_{B}^{0}\leq 1$, which is not generally true. On the other hand, the probabilistic interpretation of the two terms in $1=P(r<d)+P(r>d)$ is guaranteed for any energy-independent finite-range potential, and any possible value of $E_{B}$.

\section{Conclusions}
\label{sec:conclusions}

We have shown that the splitting into ``compositeness'' and ``elementariness'' of a weakly bound s-wave state in an energy-independent short-range potential is very close to the splitting of unity into the probabilities $P(r>d)$ and $P(r<d)$, the probabilities to find the particle outside/inside the range of the potential (or, in the overall picture of two particles in their c.m. frame, to find the particles separated from another by a distance greater/smaller than the range of their mutual interaction). This once again supports the concept of ``compositeness'' as a useful assessment of the composite nature of a given bound state. From the above observations, one might wish to replace the name ``elementariness'' by something like ``confinedness''. In QCD, of course, one expects a confining potential exactly for the interaction between the quarks, and in this respect, relating our $d$ to a scale like the confinement radius, the usual interpretations in terms of ``hadronic molecules'' vs. ``quark-antiquark states'' etc. could even be justified in the present simple framework. For the case of infinite-range potentials, let alone for the case of fully relativistic quantum field theory, we believe that such a simple interpretation as proposed here will not be available, but it might be worthwhile to search for appropriate generalizations.

\section*{Acknowledgement}
This work was supported by the Czech Science Foundation GACR grant 19-19640S.

\end{document}